\documentclass[12pt]{article}

\usepackage[utf8]{inputenc}
\usepackage[T1]{fontenc}
\usepackage{amsmath,amssymb}
\usepackage{graphicx}
\usepackage{booktabs}
\usepackage{float}
\usepackage{array}
\usepackage{tabularx}
\usepackage[margin=1in]{geometry}
\usepackage{titling}
\usepackage{hyperref}
\usepackage[super,comma,sort&compress]{natbib}
\usepackage{etoolbox}
\makeatletter
\renewcommand\@biblabel[1]{\makebox[2em][r]{[#1]}}
\makeatother

\title{Scaling Interferometry to the Multi-Petawatt Regime}

\author{\small
  M.~L.~Klebonas$^{1}$, H.~J.~Quevedo$^{1}$, C.~I.~Hojbota$^{1}$, M.~Spinks$^{1}$, K.~Riordan,$^{1}$
  \\ \small R.~Nedbailo$^{1}$, M.~Downer$^{1}$, H.~G.~Rinderknecht$^{2}$, and O.~Z.~Labun$^{1}$ \\[1ex]
\small  $^{1}$The University of Texas at Austin \\
 \small $^{2}$Laboratory for Laser Energetics, University of Rochester
}
\date{}
\begin{document}

\maketitle
\vspace{-2.5em}
\begin{abstract}
Pre-plasma conditions strongly influence laser-plasma interactions in the multi-petawatt (MPW) regime, increasing the need for reliable early-time plasma evolution diagnostics. Among available pre-plasma diagnostics, interferometry remains the most direct method for measuring the spatially resolved electron density of pre-formed plasmas. However, its implementation becomes increasingly challenging at MPW scale due to steep density gradients, phase-recovery difficulties, strong electromagnetic pulses (EMP), debris accumulation, and high-repetition-rate operation. Compounding these technical challenges, many large-scale facilities lack permanent probe-line architecture and trained diagnostic support, reducing experimental reproducibility and consuming limited beamtime. Future MPW facilities should standardize probe-line architecture, adopt off-frequency probing strategies, improve phase-recovery methods for non-symmetric plasmas, integrate emerging real-time analysis capabilities, and engineer diagnostic systems resilient to EMP and high-repetition-rate environments. These advances will enable the user community to reliably characterize pre-plasma formation and laser-plasma dynamics at next-generation MPW facilities.
\end{abstract}

\section{Problem Statement and Context}

As next-generation MPW laser systems come online, pre-plasma characterization becomes increasingly critical. At intensities approaching and exceeding $10^{23}$~W/cm$^{2}$, laser pre-pulse with energy of $10^{-6}$--$10^{-8}$ of the main pulse may be deposited nanoseconds to picoseconds before peak irradiance. Such pre-pulses can make early-time plasma expansion, ionization dynamics, and density scale length strongly govern laser coupling efficiency, the onset of relativistic transparency, and energetic particle and radiation generation~\cite{Hornung2021,Flacco2008}. The resulting pre-plasma density scale lengths can range from sub-micron to tens of microns, depending on target type and contrast conditions, and small variations in these parameters can qualitatively alter the physics of the main-pulse interaction.

Across the diagnostic technology landscape, no single tool addresses all aspects of pre-plasma evolution. Shadowgraphy provides qualitative density channel morphology; Thomson scattering offers local electron temperature and density with high spectral resolution but limited spatial coverage; X-ray diagnostics probe compressed or heated solid-density regions. Interferometry remains uniquely suited for providing quantitative, spatially resolved electron density maps of the expanding pre-plasma~\cite{Quevedo2016,Antici2010}, and is directly relevant to relativistic photon-source experiments, laser-wake dissipation measurements, and broader studies of energy coupling and secondary-source generation~\cite{Rinderknecht2021,Zgadzaj2020}. However, the MPW environment simultaneously stresses interferometric diagnostics across multiple fronts: rapidly evolving density gradients distort phase recovery, steep gradients refract probe beams away from the detector, strong EMP disrupts cameras and electronics, and high-repetition-rate operation introduces thermal loading, debris accumulation, and large data volumes requiring automated processing. Many facilities also lack permanent probe-line infrastructure, forcing users to construct temporary systems under tight scheduling constraints, which reduces reproducibility and diverting beamtime from science. These challenges motivate treating interferometric diagnostics not as an individual user responsibility, but as core facility infrastructure requiring dedicated design, personnel, and long-term development.

\section{Recommendations}

\subsection{Phase Recovery and Probe Design}

Reliable phase recovery in the presence of steep, rapidly evolving density gradients is a fundamental requirement at MPW scale. Standard assumptions, such as cylindrical symmetry for Abel inversion, increasingly break down as plasma conditions become more turbulent and asymmetric~\cite{Quevedo2016,Antici2010,Gizzi2006}. The plasma refractive index is given by $ n \approx \sqrt{(1 - \frac{n_e}{n_{cr}}})$ which scales as $n_{cr} \propto \lambda^{-2}$. Consequently, shorter-wavelength probe lines can propagate through higher-density plasma regions with reduced refraction and less severe cutoff effects~\cite{Flacco2008,Zgadzaj2020}. Optimized probing geometries and phase-retrieval methods robust to symmetry violations should be prioritized. These include Fourier-transform-based phase extraction and emerging approaches which do not require symmetry assumptions. Machine-learning-assisted fringe analysis, which is analogous to ML pipelines demonstrated for high-repetition-rate Thomson scattering~\cite{Eisenbach2025}, offers a promising route to robust phase retrieval from degraded or low-contrast interferograms. Single-shot multi-frame interferometry, which captures temporally resolved snapshots within a single laser shot, is a particularly valuable emerging capability for resolving rapid density evolution without requiring shot-to-shot reproducibility. Where interferometry cannot fully resolve fringe degradation, complementary diagnostics such as Thomson scattering or shadowgraphy may be necessary, though at the cost of additional alignment complexity and personnel requirements.

For signal isolation, frequency-shifted probes generated via harmonic conversion of a main laser pick-off are a current standard but often require supplementary scatter rejection techniques. A more flexible long-term direction is an independent synchronized probe laser at a distinct wavelength~\cite{Bernert2022}, offering greater spectral separation and reduced sensitivity to upstream main-laser variations (see Table~\ref{tab:probe-strategies}). The principal challenge then shifts from scatter isolation to synchronization: early-time plasma dynamics are sensitive to timing errors on the order of picoseconds~\cite{Hornung2024}, placing strict requirements on long-term stability and shot-to-shot reproducibility.

\begin{table}[H]
  \centering
  \footnotesize
  \renewcommand{\arraystretch}{1.4}
  \caption{Trade-offs between the three primary probe strategies available at MPW scale.}
  \label{tab:probe-strategies}
  \begin{tabularx}{\textwidth}{@{}>{\centering\arraybackslash}p{2.4cm}>{\centering\arraybackslash}c>{\centering\arraybackslash}c>{\centering\arraybackslash}c>{\centering\arraybackslash}X@{}}
    \toprule
    \textbf{Probe Strategy} & \textbf{Signal Isolation} & \textbf{Sync.\ Complexity} & \textbf{MPW Scalability} & \textbf{Note} \\
    \midrule
    Same-frequency pick-off & Poor & Low & Limited & Scatter contamination severe at MPW \\
    Harmonic-converted pick-off & Good & Low & Moderate & Requires spatial filtering, ND attenuation, or cross-polarization \\
    Independent synchronized laser & Excellent & High & High (if stabilized) & Picosecond timing precision required; demonstrated at PW scale~\cite{Bernert2022} \\
    \bottomrule
  \end{tabularx}
\end{table}

\subsection{Facility Infrastructure Standardization}

Experimental efficiency at large-scale facilities is significantly limited by the absence of permanent, standardized probe-line infrastructure. Users frequently face extended setup time to construct temporary probe lines or forego interferometric measurements entirely under strict scheduling constraints. This produces inconsistent diagnostic configurations across campaigns and reduces reproducibility in ways that undermine the scientific value of limited beamtime. Experience at existing petawatt-class facilities highlights that probe-line design is non-trivial: port placement, beam transport path length, and depth-of-focus requirements are dependent on chamber geometry in ways that must be addressed early in facility design.

Probe-line architecture should be incorporated as a baseline component of every new MPW laser system and major facility upgrade. The OMEGA-EP optical diagnostic suite provides an example of a facility-standard $4\omega$ probe diagnostic~\cite{Froula2012}, which has proven particularly successful for shadowgraphy and angular filter refractometry~\cite{Heuer2024}. Dedicated probe beam transport paths, designated target chamber ports for interferometric access, and careful optical layout during facility design can dramatically reduce alignment complexity and improve cross-campaign reproducibility. Equally important is the presence of on-site personnel with probe-line expertise who can support visiting users, streamline troubleshooting timelines, and ensure interferometric measurements remain technically feasible across diverse experimental configurations. Establishing shared design standards across MPW facilities would further improve reproducibility and create a foundation for cross-facility comparison of results.

\subsection{EMP, Debris, and High-Repetition-Rate Operation}

The MPW environment presents four coupled environmental stressors. First, strong EMP generated during high-intensity shots, with electric field strengths that can reach tens of kV/m at meter-scale distances from the target~\cite{Nelissen2020}, can disrupt cameras, timing electronics, and data acquisition hardware, directly compromising data capture. Interferometric cameras and associated electronics should be protected through proper shielding, grounding, physical isolation, and remote placement where practical. Systematic characterization of EMP environments at MPW facilities, and development of community-wide hardening standards for diagnostic electronics, would benefit reproducibility across institutions.

Second, debris accumulation on exposed optics reduces probe transmission and fringe visibility, particularly at the higher shot rates envisioned for next-generation systems. Protective strategies, including debris shielding, retractable optic mounts, and real-time probe transmission monitoring, should be incorporated into probe-line designs.

Third, high-repetition-rate operation introduces thermal loading on transmissive optics, producing wavefront distortion and thermal lensing that degrade interferogram quality~\cite{Heuer2022,Fourmaux2009}. Active thermal management techniques and periodic optic replacement should be anticipated when operating within this regime.

Fourth, the large data volumes produced at high repetition rates require automated interferogram processing and real-time or near-real-time phase retrieval pipelines to maintain diagnostic utility without manual bottlenecks. Data analysis architecture must account for the large volumes produced at high repetition rates, requiring on-site computing infrastructure and standardized data formats that support efficient archiving and cross-experiment analysis.

\section{Key Challenges and Future Directions}

\textbf{Phase recovery in asymmetric plasmas:} Standard deconvolution techniques, such as Abel inversion, fail for non-cylindrically symmetric or turbulent plasmas; true 3D reconstruction may require multiple probe axes and regularization methods.

\vspace{0.8em}\noindent\textbf{Synchronization of independent probe lasers:} Achieving and maintaining picosecond-level timing precision between an independent probe laser and the MPW master clock remains an open engineering challenge, particularly as facility repetition rates increase.

\vspace{0.8em}\noindent\textbf{Automated, real-time fringe analysis:} Processing interferograms at high repetition rates requires fast, reliable algorithms, including ML-based approaches, capable of handling degraded or low-quality fringes without user intervention.

\vspace{0.8em}\noindent\textbf{EMP characterization and hardening standards:} Systematic EMP field mapping at MPW facilities and development of community-wide electronics hardening standards would benefit diagnostic reproducibility across institutions.

\vspace{0.8em}\noindent\textbf{Facility-level standardization and open data:} Shared probe-line design standards and standardized data formats across MPW facilities would improve reproducibility, reduce user onboarding timeline, and support cross-facility physics comparisons.

\clearpage
\begin{raggedright}

\end{raggedright}

\end{document}